\documentclass[]{iopart}
\usepackage{graphics}
\usepackage{graphicx}
\usepackage{amsfonts}
\usepackage{amssymb}
\usepackage{cite}
\usepackage{xcolor}
\usepackage{amsthm}
\usepackage{amsmath}
\usepackage{amscd}
\usepackage[most]{tcolorbox} 

\theoremstyle{plain}

\def\beqa{\begin{eqnarray}}
\def\eeqa{\end{eqnarray}}

\def\hw{\hbar \omega}
\def\hw4{ \frac {\hbar \omega}{4}}

\def\uni{{\bf i}}
\def\re{{\rm {e}}}

\def\b0{b_0}

\def\dag{^\dagger}
\def\pt{{\mathcal{PT}}}
\def\nonn{\nonumber \\}
\def\uz{U_z(sl(2, \mathbb R))}
\newcommand{\RR}{\mathbb{R}}

\newcommand{\NN}{\mathbb{N}}

\begin{document}

\title{Non-standard quantum algebras and infinite-dimensional
{\cal{PT}}-symmetric systems}

\author{Angel Ballesteros$^{1}$, Romina Ram\'\i rez$^{2}$ and Marta Reboiro$^{3}$ }

\address{{\small\it $^{1}$} Departamento de F\'{i}sica, Universidad de Burgos, Burgos, Spain}
\address{{\small\it $^{2}$} IAM, CONICET-CeMaLP, University of La Plata, Argentina}
\address{{\small\it $^{3}$}IFLP, CONICET-Department of Physics, University of La Plata, Argentina}

\ead{angelb@ubu.es}
\ead{romina@mate.unlp.edu.ar}
\ead{reboiro@fisica.unlp.edu.ar}

\vspace{10pt}
\begin{indented}
\item[] \today
\end{indented}

\begin{abstract}
In this work, we introduce a 
$\pt$-symmetric infinite-dimensional representation of the 
$\uz$ Hopf algebra, and we analyse a multiparametric family of Hamiltonians constructed from such representation of the generators of this non-standard quantum algebra. It is shown that all these Hamiltonians can be mapped to equivalent systems endowed with a position-dependent mass. From the latter presentation, it is shown how appropriate point canonical transformations can be further defined in order to transform them into Hamiltonians with constant mass over suitable domains. By following this approach, the bound-state spectrum and the corresponding eigenfunctions of the initial $\pt$-symmetric Hamiltonians can be determined. It is worth stressing that a relevant feature of some of the new $\uz$ systems here presented is found to be their connection with double-well and P\"oschl-Teller potentials. In fact, as an application we present a particular Hamiltonian that can be expressed as an effective double-well trigonometric potential, which is commonly used to model several relevant systems in molecular physics.
\end{abstract}
\vspace{2pc}

\section{Introduction.}\label{intro}

$\pt$-symmetric Hamiltonians have been proven to play an essential role in describing new dynamical features of quantum systems both theoretically \cite{theo,ueda,ref1,ref3} and experimentally \cite{exp,exp1}. In particular, $\pt$-symmetric systems undergo a so-called dynamical-phase transition \cite{hpatw,hpatw2}, which occurs in the space of parameters of the system. The so-called `exact symmetry phase' is characterized by real eigenvalues and their eigenfunctions are $\pt$-symmetric, while the spectrum of the `broken symmetry phase' contains pairs of complex-conjugate eigenvalues and their eigenfunctions are non-longer $\pt$-symmetric. The boundary between both regions is formed by Exceptional Points (EPs), where two or more eigenvalues are coalescent, together with their corresponding eigenvectors (see, for instance~
\cite{PTs,ali1,pt1,pt2,pt3,pt4,pt5,pt6,pt7,pt8,ref2}  and references therein).

On the other hand, we recall that Position-Dependent-Mass (PDM) systems have received outstanding attention due to their applicability in different contexts, which include condensed matter physics \cite{cmatter1,cmatter2}, particle physics \cite{particle1,particle2}, astrophysics \cite{astro1,astro2} and they have been used to model systems with variable density, such as semiconductors, and to study the behaviour of particles in gravitational fields \cite{grav1,grav2}. Also, in~\cite{pragapdm2020,pragapdm2022} a PDM approach is combined with different potentials in double heterostructures to study energy spectra, focusing on boundary conditions and estimating the spectra, with applications in double parabolic quantum wells for lasers. Finally, PDM Hamiltonians have been studied in the context of fractional calculus \cite{fr1,fr2}. 

In general, the formulation of a PDM  problem can be often combined with other techniques to achieve either exact solutions or good estimates. For instance, in \cite{quesne} a method is introduced that merges deformed shape invariance with point canonical transformation  (PCT) methods \cite{PCT,PCTcomplex} to generate exactly solvable PDM potentials, facilitating the derivation of bound-state energy spectra and wave functions in the Schr\"{o}dinger equation. 

In a recent work~\cite{nos24}, we have established the connection between $\pt$-symmetric Hamiltonians and non-standard Hopf algebras~\cite{uzfirst,BH1,BH2} through the construction of a $\pt$-symmetric finite-dimensional representation of the non-standard $\uz$ quantum algebra. In this way, $\pt$-symmetric Hamiltonians with non-standard quantum algebra symmetry were introduced, giving rise to a variety of new exactly solvable models. We recall that the non-standard (or Jordanian) quantum deformation of the Lie algebra $sl(2, \mathbb R)$, called $U_{z}(sl(2, \mathbb R))$, is obtained by deforming both the commutation relations and the coproduct map of the universal enveloping algebra of the classical Lie algebra $sl(2, \RR)$ in terms of a `quantum' deformation parameter $z$. This deformation is related to a specific constant solution of the Classical Yang-Baxter equation on $sl(2, \RR)$ and its representation theory has been fully studied (see~\cite{uzfirst,BH1,BH2,BH3,BH4} and references therein). 

The aim of this paper is to present a natural extension of~\cite{nos24} to the case of $\pt$-symmetric Hamiltonians constructed from an infinite-dimensional irreducible representation of the $\uz$ quantum algebra. We will show that such a generalization provides an interesting link between quantum algebras and Position-Dependent-Mass (PDM) systems. Moreover, the use of point canonical transformations (PCT) will come into play as an essential ingredient for obtaining the solutions of the new $\pt$-symmetric quantum systems here presented.

The paper is organised as follows.
Section \ref{form} overviews the key concepts to be explored. It delves into the mathematical structures of the non-standard Hopf algebra $\uz$ and the construction of a $\pt$-symmetric infinite-dimensional representation of this quantum algebra. Afterwards we introduce a multiparametric family of $\pt$-symmetric Hamiltonians written in terms of the generators of $\uz$, which will be presented as PDM Hamiltonians by making use of similarity transformations. Then, it is shown how by applying appropriate PCTs, these systems will be reformulated as Hamiltonians with effective constant mass over  specific domains. Section \ref{results} studies specific examples of Hamiltonians belonging to the family that has been previously introduced, and the solutions to these systems are obtained by making use of the procedure introduced in Section \ref{form}. All of them can be written in terms of a constant mass Schr\"odinger equations, and their bound spectrum together with their corresponding eigenfunctions will be presented. The connection among these systems and the ones obtained by considering the finite dimensional representations of  $\uz$ studied in~\cite{nos24} is analysed. Also, the $z\to \infty$ limit of this quantum deformation can be shown to be meaningful in the context of one of the systems here presented. As an application dealing with realistic problems, in Section \ref{appli} we make use of our formalism to propose and solve an effective Hamiltonian related to a family of Double Well Trigonometric (DWT) potentials amenable to model different physical systems \cite{sitnitsky2017a,sitnitsky2017b,sitnitsky2018,sitnitsky2019,
sitnitsky2020,sitnitsky2021,sitnitsky2023}, e.g.~the inversion of the nitrogen atom in ammonia (NH$_3$) \cite{H5O2} or the energy levels for the hydrogen bound in the KHCO$_3$ cristal \cite{H5O2}. Finally, in Section \ref{con} we present our conclusions.

\section{Formalism} \label{form}

In this section, we shall review the fundamental aspects of the non-standard Hopf algebra $\uz$. We shall introduce a particular infinite-dimensional $\pt$-symmetric irreducible representation for this quantum algebra. After that, we shall briefly present the essentials of the formalism of the PDM Hamiltonians and the PCT technique, that we shall apply throughout the paper in order to obtain the spectrum and the eigenfunction of different families of $\pt$-symmetric Hamiltonians.

\subsection{Non-standard $\uz$ algebra}

As it is well known, the generators of the $sl(2,\RR)$ algebra obey the commutation relations:
\begin{equation}\label{comm0}
[l_{0},l_{\pm}]=\pm 2 l_{\pm} \ \ \ \ \ \ [l_{+},l_{-}]=  l_{0}.
\end{equation}

The $\uz$ algebra is a Hopf algebra deformation of the universal enveloping algebra of the $sl(2,\RR)$ Lie algebra. The commutation relations for the generators of the non-standard Hopf algebra $U_{z}(sl(2, \RR))$, hereafter named as $\{j_{0}^{(z)},j_{\pm}^{(z)}\}$, read 
\begin{equation}\label{commz}
[j_{0}^{(z)},j_{+}^{(z)}]= \tfrac{e^{2 z j_{+}^{(z)}}-1}{z}, \ \ \ \ [j_{0}^{(z)},j_{-}^{(z)}] = -2 j_{-}^{(z)}+z (j_{0}^{(z)})^{2}, \ \ \ \ [j_{+}^{(z)},j_{-}^{(z)}]= j_{0}^{(z)}
\end{equation}
being $z$ a real parameter. The reader is kindly referred to \cite{uzfirst,BH1,BH2,BH3,BH4} and references therein for further information.

We can represent the action of the operators~\eqref{commz} onto the space of states of a quantum harmonic oscillator by using the bosonic realization proposed in \cite{BH1,BH2}, where the boson operators   $a$ and $a^\dagger$ obey the usual commutation relation $[a,a^\dagger]=1$. Such a deformed Gelfan'd-Dyson realisation reads 
\begin{eqnarray}
    {j}_{+}^{(z)} & = &   a^{\dag} , \nonumber \\
    {j}_{0}^{(z)} & = &
 \tfrac{\re^{ 2 z a^{\dag}} - 1}{z} a -\lambda \tfrac{ \re^{2  z a^{\dag}} + 1}{2}, \nonumber \\
    {j}_{-}^{(z)} &=& - \tfrac{\re^{2 z a^{\dag}} -1}{2  z}a^{2}+\lambda \tfrac {\re^{2 z a^{\dag}} + 1}{2} a -z \lambda^{2}\tfrac{\re^{2 z a^{\dag}} -1}{8},
\end{eqnarray}
where $\lambda\in\mathbb{R}$ labels different representations. In fact, the Casimir operator of the quantum algebra~\eqref{commz} is
\begin{flalign}
C= \frac 12 j_0^{(z)}\re^{-2 z j_+^{(z)}}j_0^{(z)} + \tfrac{1-\re^{-2 z {j}_{+}^{(z)}}}{2 z} j_-^{(z)} + j_-^{(z)} \tfrac{1-\re^{-2 z {j}_{+}^{(z)}}}{2z}+ \re^{-2 z {j}_{+}^{(z)}}-1,
\end{flalign}
and its eigenvalue is given by $\tfrac 12 \lambda (\lambda + 2 )$. As it was shown in \cite{BH3}, for generic values of $\lambda$ this representation is infinite-dimensional and irreducible, but when $\lambda \in {\mathbb Z}^-$, the representation becomes reducible and gives rise to finite-dimensional irreducible modules with dimension $d=|\lambda-1|$. In this work we consider only the case $\lambda >0$.

Furthermore, we can realise the action of $a^{\dagger}$ and $a$ on a function $f(x)$ by the operators $x$ and $ \frac {d}{dx} $, respectively. Thus, in this differential realisation 
\begin{eqnarray}
    {j}_{+}^{(z)} & = & x , \nonumber \\
    {j}_{0}^{(z)} & = &
 \tfrac{\re^{ 2 z x} - 1}{z}  \frac{d}{dx} -\lambda \tfrac{ \re^{2  z x} + 1}{2}, \nonumber \\
    {j}_{-}^{(z)} &=& - \tfrac{\re^{2 z x} -1}{2  z}  \frac{d^{2}}{dx^{2}}+\lambda \tfrac {\re^{2 z x} + 1}{2}  \frac{d}{dx} -z \lambda^{2}\tfrac{\re^{2 z x} -1}{8}.
\label{dif}
\end{eqnarray}
Obviously, in the limit $z \rightarrow 0$, we obtain the usual differential representation for $sl(2,\RR)$ generators
\begin{eqnarray}
l_{+}&=& x\,I, \nonn
l_{0}&=& 2 x \frac{d}{dx} -\lambda\, I,\nonn
l_{-}&=& -x \frac{d^2}{dx^2}+\lambda \frac{d}{dx}\, ,
\label{sl2x}
\end{eqnarray}
where $I$ is the identity operator.

From~\cite{nos24} it is straightforward to prove that the $\pt$-symmetric realisation of the $\uz$ quantum algebra~\eqref{commz} is given by 
\begin{flalign}\label{ptdif}
& J_{+}^{(z)}= -\uni j_{+}^{(-\uni z)}=-\uni x I \nonumber \\
& J_{0}^{(z)}=  j_{0}^{(-\uni z)}=\frac{2 e^{-\uni x z}  \sin (x z)}{z}\frac{d}{dx}-\lambda  e^{-\uni x z} \cos (x z)I 
\\
& J_{-}^{(z)}=  \uni j_{-}^{(-\uni z)}=-\frac{\uni e^{-\uni x z} \sin (x z)}{z} \frac{d^{2}}{dx^{2}}+\uni \lambda  e^{-\uni x z}  \cos (x z)\frac{d}{dx}+ 
 \frac{1}{4} \uni \lambda ^2 z e^{-\uni x z} \sin (x z)I,  \nonumber  
\end{flalign}
which, in the limit $z \rightarrow 0$, leads to the usual $\pt$-symmetric representation of the $sl(2,\RR)$ algebra
\begin{eqnarray}
L_{+}&=& -\uni l_{+}= -\uni x  I, \nonn 
L_{0}&=& l_{0}= 2 x \frac{d}{dx}-\lambda  I, \nonn 
L_{-}&=& \uni l_{-}= \uni \lambda  \frac{d}{dx}-\uni x \frac{d^{2}}{dx^{2}}. 
\label{sl2pt}
\end{eqnarray}

In this paper we shall consider with the following family of Hamiltonians written in terms of the generators of the $\uz$ quantum algebra:
\begin{equation}\label{hamil}
H=\mu_-(J_{+}^{(z)} ) \ \ J_{-}^{(z)} + \mu_0(J_{+}^{(z)} ) \ \ J_{0}^{(z)}  + \mu_+(J_{+}^{(z)} ),
\end{equation}
where $\mu_0(J_+^{(z)} )$ and $\mu_\pm(J_+^{(z)} )$ are three smooth functions of the $J_{+}^{(z)}$ generator. By construction, any Hamiltonian of the form \eqref{hamil} is PT -symmetric and is therefore endowed with all the associated properties of this kind of systems, despite $H$ will be (in general) non-Hermitian.  PT-symmetric operators may possess entirely real spectra when PT symmetry is unbroken, while in the broken phase, the spectrum typically includes complex conjugate eigenvalue pairs and exhibits Exceptional Points in the border of both regions. At the Exceptional Points two o more eigenvalues and their corresponding eigenfunctions are coalescent \cite{Swanson}. 

\subsection{PDM Hamiltonians and the PCT method}

A keystone for the results presented in this paper is the fact that we can prove that any generic Hamiltonian $H$  can be transformed into a PDM Hamiltonian, $h$, through a similarity transformation $\Gamma(x)$, namely
\begin{equation}
h  =\Gamma(x)H \Gamma^{-1}(x)= -\frac{1}{2}\frac{d}{dx}\left(\frac{1}{m(x)}\frac{d}{dx} \right)+V(x).
\label{hpdm0}
\end{equation}

Note that the PMD Hamiltonian $h$ of \eqref{hpdm0} is a particular case of the von Roos operator $H_R$ \cite{Roos} 
\begin{flalign}
H_R =\frac{1}{ 4} \left(
m(\hat x)^{\eta} ~\hat p~m(\hat x)^\varepsilon \hat p~m(\hat x)^{\rho}+
m(\hat x)^\rho \hat p~m(\hat x)^{\varepsilon}~\hat p~m(x)^{\eta} \right)+V(\hat x),
\end{flalign}
in the case $\eta =\rho=0$ and $\varepsilon=-1$, being $\hat x$ and $\hat p$ the usual position and momentum operators, respectively. 

Indeed, several PDM models represented by \eqref{hpdm0} are exactly solvable ones. In fact, a possible approach in order to solve PDM Schr\"odinger equations consists in using the so-called PCT method~\cite{PCT,PCTcomplex}, which transforms the initial PDM Hamiltonian into a Schr\"odinger equation with constant (effective) mass by means of a redefinition of the coordinate $x$ and its canonically conjugate momenta $p$. More explicitly, the PCT method is implemented through the following steps:

\begin{enumerate}
\item Consider $m(x)= m_0\,M(x)$ where $m_0$ will be a constant effective mass
\item Let us define 
\begin{equation}
u= w(x)= \int \sqrt{ M(x)} \,dx  \, ,
\end{equation}
and then (within a suitable domain) we have $x=F(u)=w^{-1}(u)$. Given a function $\varphi(x)$, the transformation of the differentials reads
\begin{eqnarray}
\frac{d \varphi}{dx} &=& \frac{d u}{d x} \frac{d \varphi}{du}\, ,\\
\frac{d^{2} \varphi}{dx^{2}} &=& \left(\frac{d \varphi}{du}\right) \frac{d^{2}u}{dx^{2}}+\left( \frac{du}{dx}\right)^{2} \frac{d^{2}\varphi}{du^{2}} \, .
\end{eqnarray}
\item We shall call 
\begin{equation}
W(u)=\frac{d}{du} \left(\ln (M(F(u)))^{-1/4} \right) \, .
\end{equation}
\item Let
\begin{equation}
U(u)=V(F(u))+\frac{1}{2 m_{0}} (W^{2}(u) + W'(u)).
\label{potentialU}
\end{equation}
\end{enumerate}

It turns out that by applying this PCT transformation, we can map \eqref{hpdm0} into a Schr\"odinger equation with constant mass of the form
\begin{equation}\label{eq1}
\mathfrak{h}\psi(u)=\,- \frac{1}{2 m_0} \psi''(u)+ U(u) \psi(u)= E \,  \psi(u)\, ,
\end{equation}
where the relation between $\varphi(x)$ and $\psi(u)$ is given by
\begin{equation}\label{psiphi}
\varphi(F(u))=\psi(u) e^{- \int^{u} W(t)dt}.
\end{equation}

The extension of the PCT to complex variables was developed in \cite{PCTcomplex}.



\subsection{The $sl(2,\RR)$ case}

In order to gain some explicit understanding of the previous formalism,
let us consider the following family of Hamiltonians written in terms of the generators of the undeformed algebra $sl(2,\RR)$ 
\begin{eqnarray}
H & = & \mu_-( L_{+}) L_{-}+ \mu_0( L_{+}) L_{0}+ \mu_+( L_{+}),
\end{eqnarray}
with $\mu_+(L_+)$, $\mu_+(L_+)$ and $\mu_0(L_+)$ being smooth functions of $L_+$. We recall that this family of Hamiltonians have been previously studied in relation with $\pt$-systems in \cite{fring1,fring2}. 

As an specific example of the method here proposed, we shall present results for 
\begin{eqnarray}
H & = & \mu_- L_{-}+ \mu_0 L_{0}+ \mu_+ L_{+},
\label{Hnd}
\end{eqnarray}
with $\mu_+$, $\mu_+$ and $\mu_0$ are constant parameters, and the $sl(2,\RR)$ generators are given in terms of the differential realisation~\eqref{sl2pt}.

After defining 
\begin{flalign}
\Gamma(x)= 
\left(-\uni x \right)^{-\frac{\lambda +1}{2}} \re^{ -\uni ~\tfrac{\mu_0}{\mu_-} x},
\end{flalign}
we obtain that the PDM Hamiltonian equivalent to $H$ is
\begin{equation}
h= \Gamma(x) H  \Gamma(x)^{-1}=
-\frac 1 2 
\frac {d} {d x}
\left(  \frac{-\uni}{2 \mu_- x} \frac{d}{dx} \right)
-\uni \mu_- \left( \frac{\mu_0^2+ \mu_- \mu_+}{\mu_-^2} x +\frac 1 {4 x}  (\lambda^2+1)\right)\, .
\label{nd2}
\end{equation}
The next step consists in performing the change of variable
\beqa
-\uni x= \frac 12 u^2,
\eeqa
and after applying the PCT method we finally obtain the Hamiltonian
\begin{equation}
\mathfrak{h}= -\frac 12 \mu_- \frac{d^2}{du^2}+
\frac 12 \mu_- \kappa^2  u^2+\mu_-\frac{3+4(\lambda(\lambda+2))}{8 u^2},
\label{nd2b}
\end{equation}
where 
\beqa
\kappa^2=\left( \frac {\mu_0^2+\mu_+ \mu_-}{\mu_-^2} \right).
\eeqa
Therefore, we have obtained that the non-Hermitian $\pt$-symmetric Hamiltonian $H$~\eqref{Hnd} can be transformed into a constant mass Schr\"odinger problem with a harmonic oscillator-type potential plus a $u^{-2}$ term \cite{PCTcomplex}.

In particular, for $\mu_->0$, it is well-known that the bound spectrum of ${\mathfrak h}$ of \eqref{nd2b} is given by
\beqa
E_n= \mu_- \kappa ~(2 ( n+ 1)+\lambda),
\eeqa
and the corresponding eigenfunctions are
\begin{equation}
\psi_n(u)= {\cal N}_n 
u^{ \lambda+\frac {3} {2}}~
\re^{-\frac 1 2 \kappa  u^2}~{\rm L}_n^{\lambda+1} \left( \kappa  u^2 \right),
\end{equation}
where ${\rm L}_n^{\lambda+1}(y)$ represents the generalised $(\lambda+1)-$Laguerre polynomial of order $n$.


\section{Results and discussion}\label{results}

In the following, we shall present and solve particular examples of the family of Hamiltonians given in \eqref{hamil}, namely
\begin{equation}
H=\mu_-(J_{+}^{(z)} ) \ \ J_{-}^{(z)} + \mu_0(J_{+}^{(z)} ) \ \ J_{0}^{(z)}  + \mu_+(J_{+}^{(z)} ),
\end{equation}
by making use of the formalism presented in the previous Section.


\subsection{Example 1. The case with $\mu_\pm$ and $\mu_0$ being constant functions}\label{sec32}
The non-standard deformed version of the Hamiltonian solved in Section 2.3 is 
\begin{eqnarray}
H & = & \mu_- J_{-}^{(z)}+ \mu_0 J_{0}^{(z)}+ \mu_+ J_{+}^{(z)}.
\label{hconst}
\end{eqnarray}

Its solution can be obtained by considering the operator
\beqa
\Gamma(x)= \re^{\uni x ~ \left(\frac z2  +\frac{\mu_0}{\mu_-} \right) } \left( 2 \uni \sin ( x z)\right)^{-(\lambda+1)/2},
\eeqa
and the similarity transformation:
\beqa
\mathfrak{h}& = & \Gamma(x) \, H \, \Gamma(x)^{-1},\nonn
& = & -\frac 12 \frac {{\rm d}}{{\rm d x}} \left( \frac 1 {m(x,z)} \frac {{\rm d}}{{\rm d x}} \right)+ V(x,z),
\eeqa
where
\beqa
m(x,z) & = & \frac z {t(x,z)},\nonn
V(x,z) & = & \frac 12 \mu_- z \left( 1+ \frac {(\lambda+1)^2} {t(x,z)}   \right)
-\frac {\mu_0^2} {2 \mu_- z } t(x,z)-\uni x \mu_+,
\eeqa
where we have defined:
\beqa
t(x,z)=1-\re^{-2 \uni x z}.
\label{txz}
\eeqa

Now, by following the PCT method, if we introduce the change of variables:
\beqa
-\uni x = \ln \left( \sec(u \sqrt{z})^2 \right)/(2 z),
\eeqa
the following Schr\"odinger equation with constant effective mass is obtained
\beqa
\mathfrak{h}\, \phi(u)=  -\frac 1 2 \mu_- \frac {{\rm d }^2\phi(u)}{{\rm d}u^2}+ U (u,z)\phi(u)= \varepsilon \phi(u)\, ,
\eeqa
where the potential is given by
\beqa
U(u,z) 
& =& \frac 1 2 \mu_- z ~\beta (\beta-1) \sec(u \sqrt{z})^2+\frac{1}{2} \mu_- z~\alpha (\alpha-1) \csc(u \sqrt{z} )^2 + \nonn
& & \frac{1}{2 z} \mu_+ ~ \ln \left(\sec(u \sqrt{z} )^2\right)-\frac{1}{2 z} \frac{\mu_0^2}{\mu_-},
\eeqa
and, as expected, in the limit $z\rightarrow 0$, we recover \eqref{nd2}.

We can introduce a new variable
\beqa
y = u \sqrt{ z}, 
\eeqa
so that 
\beqa
\mathfrak{h} \, \psi(y)=  -\frac 1 2 \mu_- z \frac {{\rm d }^2\psi(y)}{{\rm d} y^2}+ U (y,z)\phi(y)= \varepsilon \phi(y),
\eeqa
and finally, the Schr\"odinger problem takes the form
\begin{equation}
- \frac {{\rm d}^2}{{\rm d}y^2} \psi(y)+ \frac{2}{\mu_- z} ( U(y,z)-\varepsilon)\psi(y)=0, 
\end{equation}
with
\begin{equation}
 \frac{ 2 U(y,z)}{\mu_- z} =
  \beta (\beta-1) \sec(y)^2+ \alpha (\alpha-1) \csc(y )^2 + 
\frac{\mu_+}{ \mu_- z^2} \ln \left(\sec(y )^2\right)-\frac{\mu_0^2}{\mu_-^2 z^2},
\label{hex2}
\end{equation}
In the previous expression 
\beqa
\alpha (\alpha -1) & = &  \frac{3}{4}+\lambda(\lambda+2) \nonn
\beta (\beta -1) & = &  \frac{ \mu_0^{2}}{\mu_{-}^{2} z^{2}}- \frac{1}{4}.
\eeqa

The trigonometric P\"oschl-Teller potential of \eqref{hex2} supports bound states when $\alpha (\alpha -1)$ and $\beta (\beta -1)$ take positive values. Moreover, to ensure the existence of bound states with square-integrable eigenfunctions, $\beta$ and $\alpha$ must satisfy $ \beta > 1 $, 
$\alpha>1$, respectively \cite{ptgeneralizado,ptbound}. 

Let us firstly consider the case with $\mu_+=0$, which can be solved analytically.

\subsubsection{Case 1: $\mu_+=0$.\\} 

For $\beta \geq 1$, the potential $U(y,z)$ is then a trigonometric P\"oschl-Teller potential, which is periodic with period $\pi$. The eigenvalues can be written as \cite{ptgeneralizado}
\beqa
E^0_n&= &\frac{1}{2}\mu_- z (2 n +\alpha +\beta )^2-\frac{ \mu_0^2}{2 \mu_- z},\nonn
     &= & \frac{1}{2}\mu_- z (2 n+ \lambda+2 )^2+\mu_0  (2 n+ \lambda+2 ).
\label{en0}
\eeqa
with $n \geq 0$.
The corresponding eigenfunctions, which satisfy the boundary condition $|\Psi_n^0(y)| \rightarrow 0$ at $y \rightarrow \pm \pi/2$, are given by
\begin{equation}
\Psi_n^0(y) = 
{\cal N}_n (\alpha,\beta)\sin^\alpha( y)\cos^\beta( y )~ _2F_1(-n,n+\alpha+\beta,1/2+\beta,\cos^2(y ))\, ,\end{equation}
where the normalization constant ${\cal N}_n (\alpha,\beta)$ reads
\begin{flalign}
&{\cal N}_n(\alpha,\beta)= \left(2 (2 n+ \alpha+\beta)\tfrac{ \Gamma(n+\beta+ \frac 12) \Gamma(n+\alpha+\beta)}
{\Gamma(n+\alpha+\frac12 )\Gamma(n+1) \Gamma(\beta+\frac12 )^2 } \right)^{\frac 12}.
\end{flalign}

It is interesting to recall that the eigenvalues of the same Hamiltonian~\eqref{hconst} for any finite-dimensional representation of $U_{z}(sl(2, \mathbb R))$ were found in \cite{nos24} to be
\begin{flalign}
E_k^{d} =\frac 12 \mu_-z k^2 \pm k \mu_0,~ k=
\left \{
\begin{array}{lll}
2 m, &~m=0,...,(d-1)/2, &~d~\text{odd},\nonn
2 m+1, &~m=0,...,d/2, &~d~\text{even},\nonn
\end{array}
\right.
\label{ek}
\end{flalign}
being $d$ the dimension of the problem. Notice that for large values of $z$, to leading order,  $E_n^0$ of \eqref{en0} reads
\beqa
E^0_n \rightarrow \frac{1}{2}\mu_- z (2 n+  \lambda +2)^2,
\label{en0b}
\eeqa
while, in the same limit, for the leading order of the eigenvalues in finite dimension, $E_k^{d}$, we have
\begin{flalign}
E_k^{d} \rightarrow \frac 12 \mu_-z k^2,~ k=
\left \{
\begin{array}{lll}
2 m, &~m=0,...,(d-1)/2, &~d~\text{odd},\\
2 m+1, &~m=0,...,d/2, &~d~\text{even}.
\end{array}
\right. 
\end{flalign}
This means that for large values of $z$ and for large finite-dimensional dimension $d$, the leading order of the spectra for both models is the same.

\subsubsection{Case 2: $\mu_+ \neq 0$\\}

\begin{center}
\begin{figure}
\includegraphics[width=0.99\textwidth]{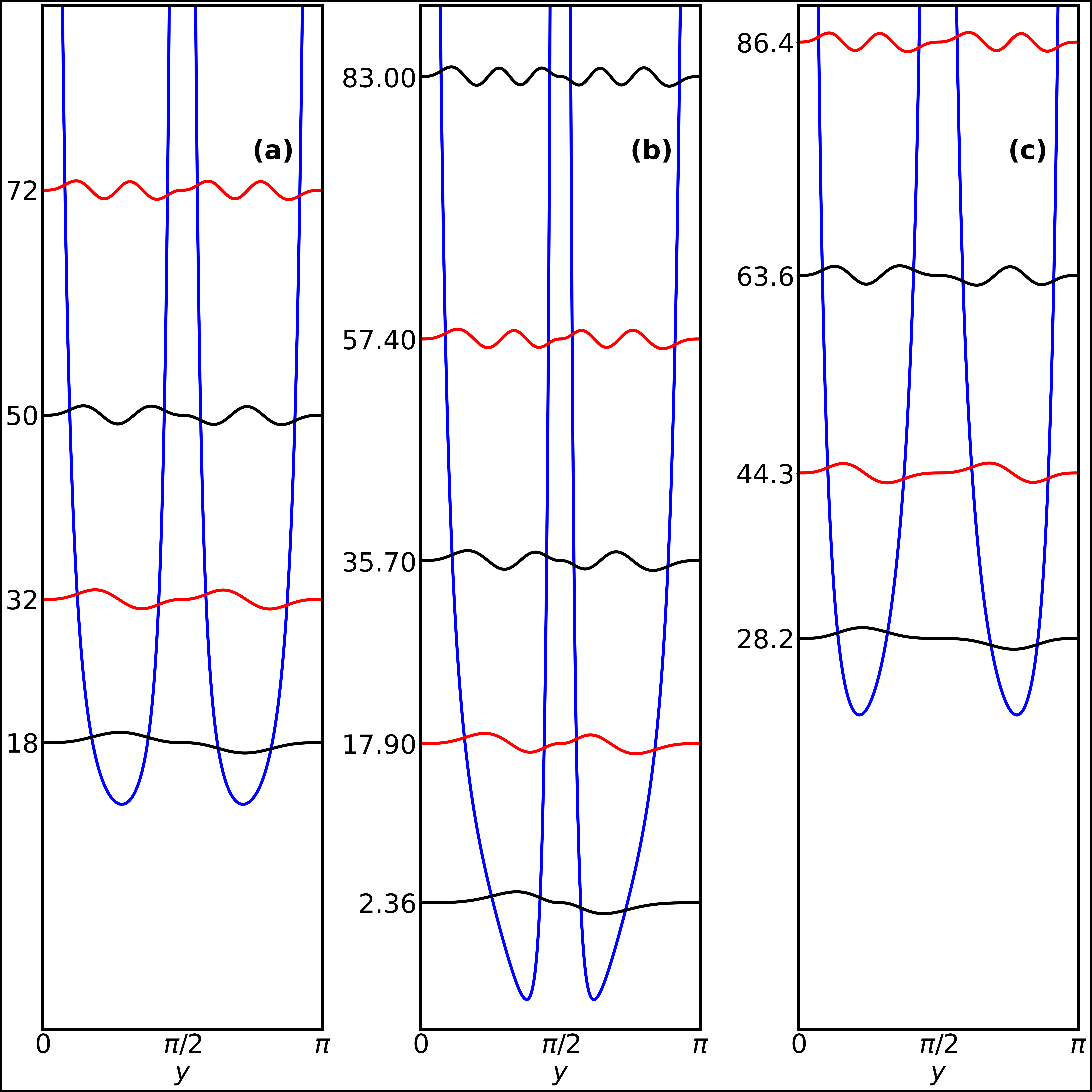}
\caption{The figure displays the behaviour of the Potential,$\frac{2}{
\mu_-z}U(y,z)$ of \eqref{hex2}, the first eigenvalues in units of $[\mu_- z]$ and the corresponding eigenfunctions, as a function of the scaled parameter $y=x \sqrt{z}$. The value of $\lambda$ and $\mu_0$ were fixed to $\lambda=2$ and $\mu_0=2$, respectively. We have taken $\mu_+=0$ for the left panel, $\mu_+=-26$ for the central panel and $\mu_+=26$ for the right panel. }
\label{fig:f1}
\end{figure}
\end{center}

\begin{center}
\begin{figure}
\includegraphics[width=0.99\textwidth]{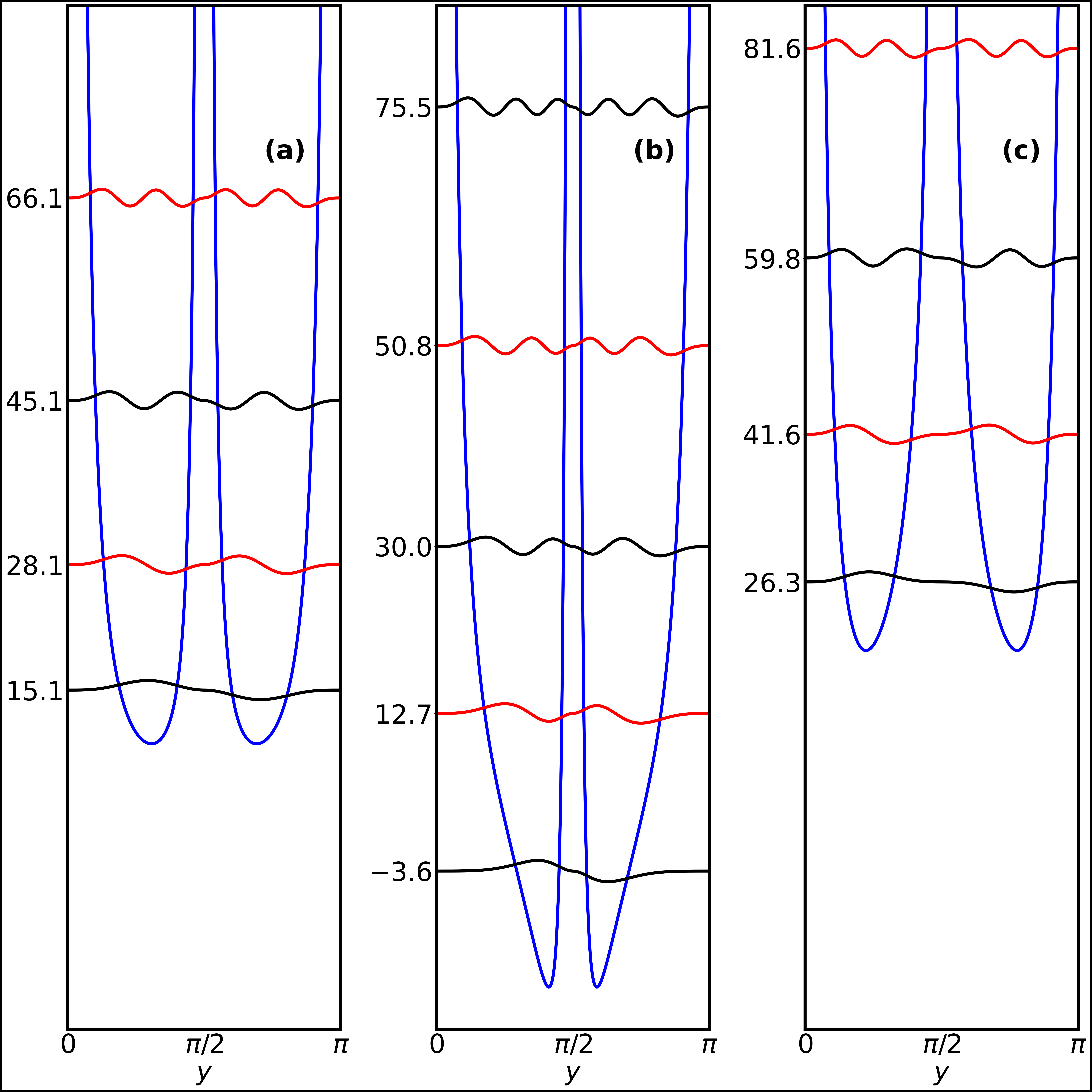}
\caption{The same as Figure \ref{fig:f1} for  $\mu_0=3/2$.}
\label{fig:f2}
\end{figure}
\end{center}

\begin{center}
\begin{figure}
\includegraphics[width=0.99\textwidth]{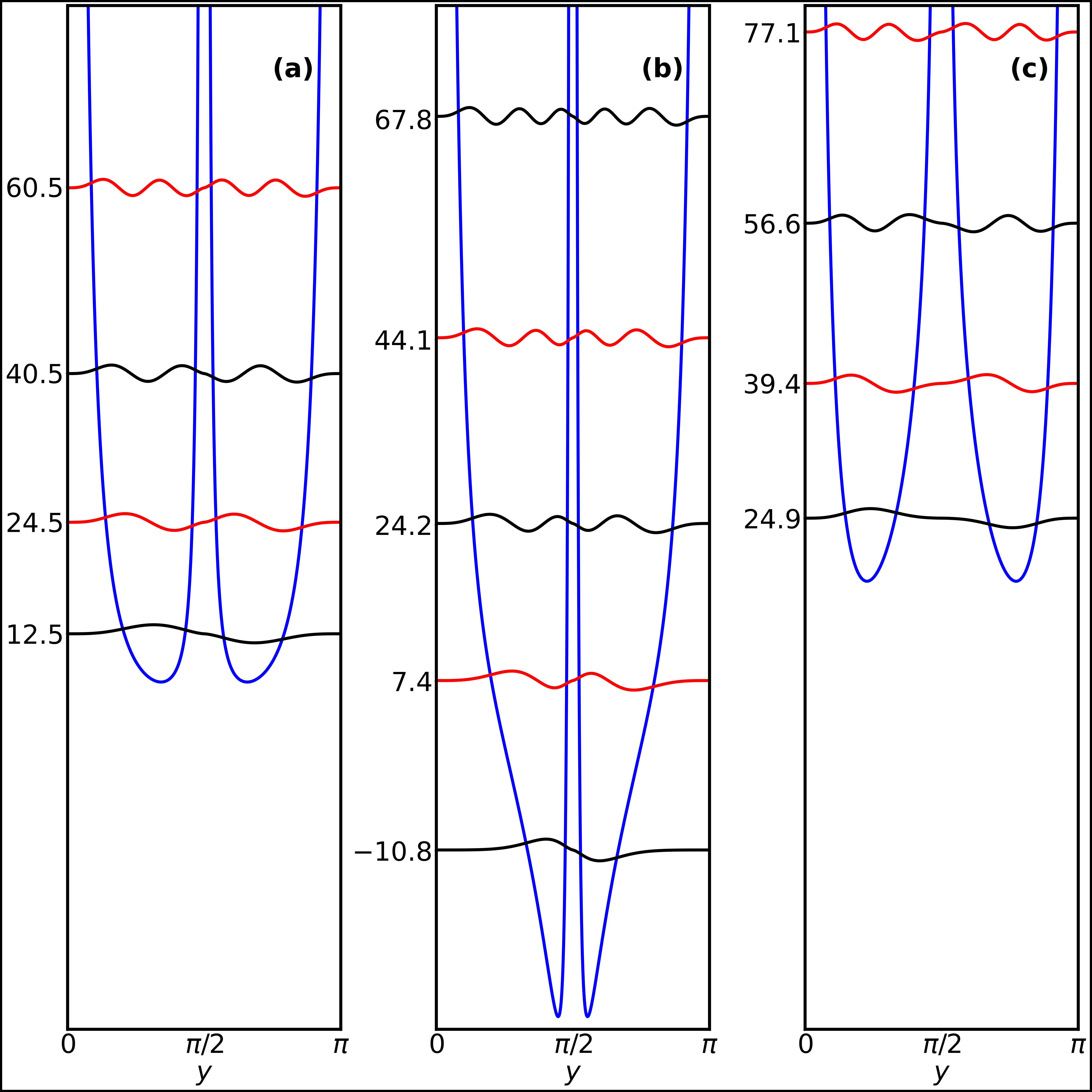}
\caption{The same as Figure \ref{fig:f1} for  $\mu_0=1$}
\label{fig:f3}
\end{figure}
\end{center}

For $\mu_+ \neq 0$, we shall study the bound spectrum and the eigenfunctions numerically. In Figure \ref{fig:f1}, we plot  the behaviour of the Potential, $\frac{2}{
\mu_-z}U(y)$, of \eqref{hex2}, its first eigenvalues in units of $[\mu_- z]$ and the corresponding eigenfunctions, as a function of the scaled parameter $y=u \sqrt{z}$. The value of $\lambda$ and $\mu_0$ were fixed to $\lambda=2$ and $\mu_0=2$, respectively. We have taken $\mu_+=0$ for the left panel, $\mu_+=-26$ for the central panel and $\mu_+=26$ for the right panel. Figure \ref{fig:f2}, shows the same information for $\mu_0=3/2$, and Figure \ref{fig:f3} for $\mu_0=1$. As it can be easily appreciated, the effect of the term with $\mu_+ \neq 0$ is to deform the original potential.

For $\mu_+ \neq 0$, we can compute the first order correction to the energy by making use of perturbation theory.  We shall firstly consider the exactly solvable model
\begin{eqnarray}
&\frac{1}{\mu_- z} \mathfrak{h}_0 \psi_n^0(y)=  -\frac 1 2 \frac {{\rm d}^2}{{\rm d y}^2} \psi_n^0(y)+ V_0(y) \psi(y)= \frac{E_n^0}{\mu_- z} \psi_n^0(y), \nonumber \\
& \qquad V_0(y)
 = \frac 1 2 \beta (\beta-1) \sec(y)^2+\frac{1}{2} \alpha (\alpha-1) \csc(y )^2 -\frac{1}{2 z^2} \frac{\mu_0^2}{\mu_-^2},\nonn
\end{eqnarray}
and afterwards the additional term coming from $\mu_+ \neq 0$ and given by:
\beqa
U(y,z) =\frac{1}{2 z^2} \frac{\mu_+}{\mu_-} \ln \left(\sec(y )^2\right).
\eeqa

A series expansion in $y=\sqrt{z} u$ shows that the term $U(y,z)$ is two orders of magnitude smaller in $z$ than the terms that give rise to the P\"oschl-Teller potential, so $U(y,z)$ 
can be taken as a perturbative term for large values of $z$.
In this limit, the correction $E_n^1$ to the unperturbed energy $E_n^0$, can be computed by standard perturbation theory methods \cite{messiah2014quantum} and reads
\begin{flalign}
E_{n}^{1} (\alpha,\beta) & = \frac{\mu_+}{2 z} \sum_{k=1}^{\infty}~\int_0^{\frac \pi 2}~\psi_n^0(x,\alpha,\beta)^* \frac {\sin(x)^{2 k}}{k} \psi_n^0(x,\alpha,\beta) {\rm{dx}}\nonn
 & = \frac{\mu_+}{2 z} \sum_{k=1}^{\infty} \sum_{m=0}^n \sum_{m'=0}^n 
2^{-k-2 n} (-1)^{-m-m'+2 n} (2 n+\alpha+\beta) v(n,\alpha,\beta,k,m,m'), \nonn
\end{flalign}
so that 
\beqa
E_n \approx E_n^0 + E_n^1.
\label{en}
\eeqa

After some algebra, the expression for $v(n,\mu,\beta,k,m,m')$ can be calculated:
\begin{flalign}
& v(n,\alpha,\beta,k,m,m') = \frac{f(n,\alpha,\beta,k,m,m')}{g(n,\alpha,\beta,k,m,m')},\nonn
& f(n,\alpha,\beta,k,m,m') = \Gamma (n+1)  \Gamma (n+\beta+\frac 12) \Gamma(n+ \alpha +\frac 12)  \Gamma (n+\alpha+\beta) \nonn
& 
~~~~~~~~ \left(\Gamma (2 n -m-m'+\beta+\frac 12) \, _2F_1(1,-k-m-m'-\alpha-\frac 12;2 n -m-m'+\beta+\frac 32;-1)+ \right.  \nonn
& 
~~~~~~~~ \left. \Gamma (k+m+m'+\alpha+\frac 12) \, 
_2F_1(1,-2 n +m+m'-\beta-\frac 12 ;k+m+m'+ \alpha+\frac 32;-1)\right)
\nonn
& 
g(n,\alpha,\beta,k,m,m') =  k ~\Gamma (m+1) \Gamma (m'+1) \Gamma (n+1-m) \Gamma(n+1-m') 
\nonn & ~~~~~~~~
\Gamma (m+\alpha+\frac 12) 
\Gamma (m'+\alpha+\frac 12) \Gamma (n-m+\beta+\frac 12) \Gamma (n-m'+\beta+\frac 12)
\nonn & ~~~~~~~~\Gamma(2 n-m-m'+\beta+ \frac 32) \Gamma (k+m+m'+\mu+\frac 32).
\end{flalign}

We recall that in the $\mu_+ \neq 0$ case, the eigenvalues of the Hamiltonian~\eqref{hconst} were also computed in~\cite{nos24}  for any finite-dimensional representation of $U_{z}(sl(2, \mathbb R))$. In the limit of large values of $z$, such eigenvalues were given by
\begin{flalign}\label{ekfd1}
E_k^{FD} =\frac 12 \mu_-z k^2 \pm  \sqrt{k^2 \mu_0^2 +k \mu_- \mu_+},~ k=
\left \{
\begin{array}{lll}
2 n, &~n=0,...,(d-1)/2, &~d~\text{odd},\\
2 n+1, &~n=0,...,d/2, &~d~\text{even},
\end{array}
\right.
\end{flalign}

\begin{center}
\begin{figure}
\includegraphics[width=0.99\textwidth]{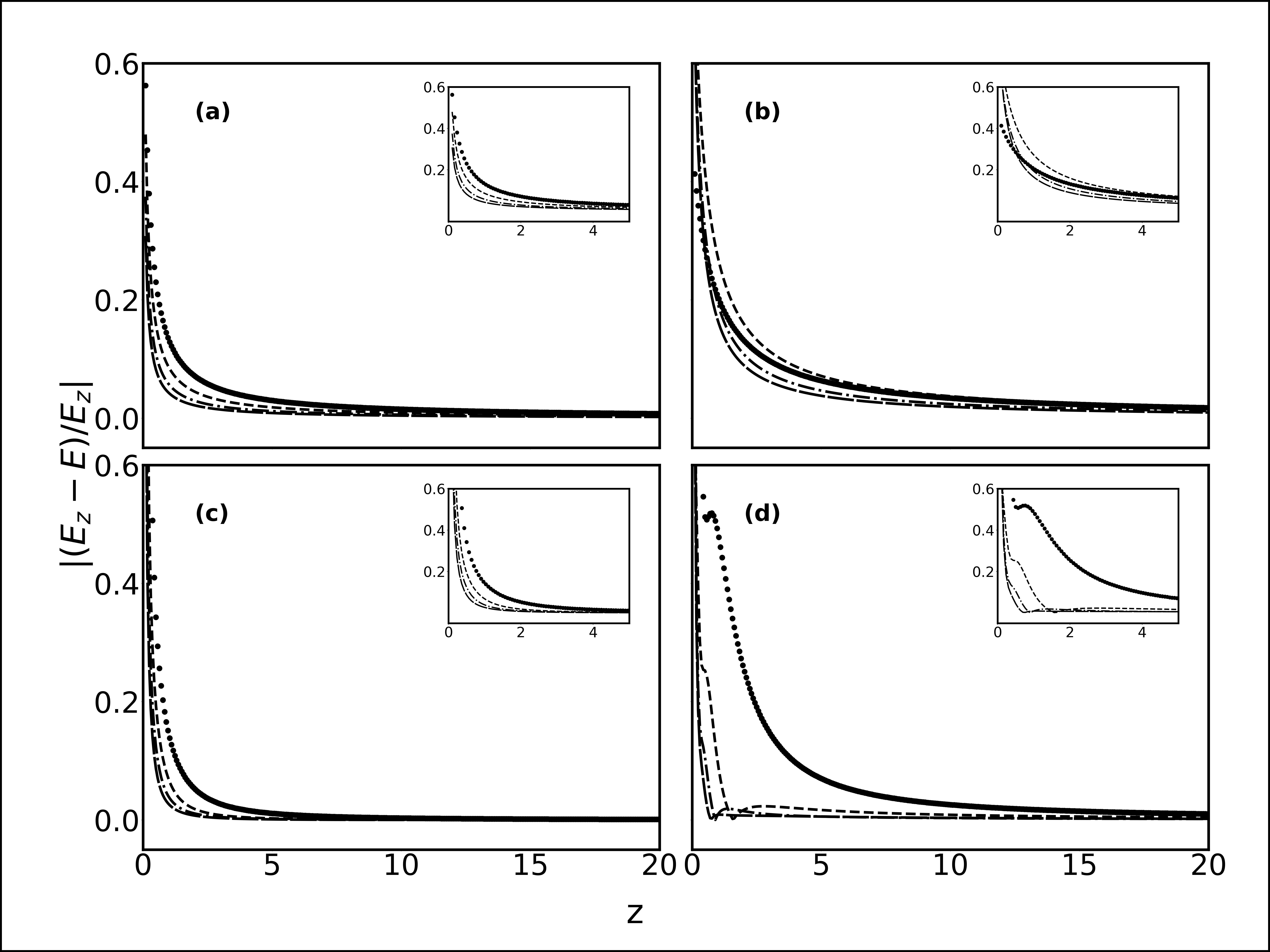}
\caption{Absolute value of the difference between $E_n$ of \eqref{en} $E_k^{FD}$ of \eqref{ekfd1} as a function of $z$. We have taken $\lambda=2$ and $\mu_-=1$. For the upper Panels, (a) and (b), $\mu_0=0$, while for the lower Panels,(c) and (d), $\mu_0=2$. Left Panels, (a) and (c), correspond to values of $\mu_+=0.5$ and the right Panels, (b) and (d), to values of $\mu_+=10$, respectively. We have plotted dotted, dashed, dashed-dotted, and long-dashed lines for $n=0,~1,~2,~3$, respectively.}
\label{fig:f4}
\end{figure}
\end{center}

In Figure \ref{fig:f4}, we display the absolute value of the difference between $E_n$ of \eqref{en} and $E_k^{FD}$ of \eqref{ekfd1} as a function of $z$. We have taken $\lambda=2$ and $\mu_-=1$. For the upper panels $\mu_0=0$, while for the lower panels, $\mu_0=2$. Left panels correspond to values of $\mu_+=0.5$ and the right panels to values of $\mu_+=10$, respectively. From Figure \ref{fig:f3} it can be seen that for large values of z the perturbative results and the eigenvalues of the finite-dimensional representation are convergent.


\subsection{Example 2}\label{sec34}

Let us now consider the Hamiltonian 
\begin{eqnarray}
H & = & \mu_- J_{-}^{(z)}+ \mu_0 J_{0}^{(z)}+ \mu_+ [J_0^{(z)},J_+^{(z)}] \, ,
\label{hj02}
\end{eqnarray}
where, from~\eqref{commz},  $[J_0^{(z)},J_+^{(z)}]=(\re^{2 z     {J}_{+}^{(z)}}-1)/z$.

We can introduce the similarity mapping $\gamma$ (see~\cite{nos24})
\begin{equation}
 \gamma= \re^{\eta J_0^{(z)}} \re^{\kappa_\pm J_+^{(z)}}, ~~~\kappa= \frac{1}{\mu_-} \left( \sqrt{\mu_0^2+ 2 \mu_+ \mu_-}-\mu_0 \right),
\end{equation}
so that 
\begin{flalign}
 H_{\eta}= \gamma H \gamma^{-1}= \mu_-e^{- 2 \eta} J_{-}^{(z)}+ z \mu_- e^{-\eta}\sinh(\eta)  \left( J_{0}^{(z)}\right)^{2} +\tau J_{0}^{(z)},
  \label{heta}
\end{flalign}
where
\beqa
\tau=\sqrt{\mu_0^2+ 2 \mu_+ \mu_-}.
\eeqa

It can be shown that the operator $H_{\eta}$ can be then mapped to a PDM Hamiltonian
\begin{equation}
h_{\eta}  =\Gamma(x)H_{\eta}\Gamma^{-1}(x)= -\frac{1}{2}\frac{d}{dx}\left(\frac{1}{m(x,z)}\frac{d}{dx} \right)+V(x,z),
\end{equation}
where
\begin{eqnarray}
&& \!\!\!\!\!\!\!\!\!\!\! \!\!\!\!\!\!\!\!\!\!\!
\Gamma(x)= \mu_- \re^{\eta} \left(\re^{-\eta} + t(x,z) \left (1- \re^{-\eta} \right )\right)^{\frac \tau {2\mu_-z}}
\left( 1-t(x,z)\right)^{\frac \lambda 4 -\frac {\tau} {2 \mu_- z}} t(x,z)^{-\frac 12 (\lambda+1)}
\nonn \\
&& \!\!\!\!\!\!\!\!\!\!\! \!\!\!\!\!\!\!\!\!\!\!
m(x,z)=\frac{z}{\mu_-} \frac {1}{\re^{-2 \eta} t(x,z) \left( 1-(1-\re^{2 \eta}) t(x,z)\right)}
\label{mseta} \nonn \\
&& \!\!\!\!\!\!\!\!\!\!\! \!\!\!\!\!\!\!\!\!\!\!
V(x,z)=
\frac{\mu_- z }{2 ~ t(x,z) \re^{2 \eta}} 
\left(~ t(x,z) ~\left (~ (3 t(x,z) -1)(\re^{2 \eta}-1)+1 ~ \right)+(\lambda +1)^2~\right)
\nonn
& & \qquad + \frac{\tau ^2~t(x,z) \re^{2 \eta} }{2 \mu_- z ~ (1-t(x,z) \re^{2 \eta})},\nonn
\label{veta}
\end{eqnarray}
where the function $t(x,z)$ is given in \eqref{txz}.

We introduce the change of variable
\begin{equation}
 -\uni x= - \frac{ \ln (1-\re^{-2 \eta} \sin(y)^2)}{2 z},~~~ y=u \sqrt{\mu_- z },
\end{equation} 
so that, we have
\begin{equation}\label{IPM}
\frac{d^2 \phi(y)}{dy^{2}}+ \frac {2 }{\mu_- z}\left( U(y,z)- \varepsilon \right) \phi(y) =0,
\end{equation}
where 
\begin{eqnarray}
\frac{ 2 U(y,z)}{\mu_- z}&=&-\frac{\tau^2}{ \mu_-^2 z^2} + \alpha (\alpha-1) \csc ^2\left(y\right)+\beta(\beta-1) \sec ^2\left( y\right),
\label{eta}
\end{eqnarray}
is again a trigonometric P\"oschl-Teller potential, with

\beqa
\alpha (\alpha - 1) & = & \frac{3}{4}+ \lambda (2+ \lambda)\\
\beta (\beta -1) & = & \frac{\mu_{0}^{2}+ 2  \mu_{0}\mu_{-}}{\mu_{-}^{2}z^{2}}- \frac{1}{4}.
\eeqa
As pointed before, $\alpha (\alpha - 1)>0$ and $\beta (\beta - 1)>0$ to obtain the bound states of the Hamiltonian of \eqref{eta}. Also, $\alpha>1$ and $\beta>1$ in order to guarantee square-integrable eigenfunctions. 
Notice that the limit $\eta\rightarrow 0$ of this construction, which corresponds to the Hamiltonian
\begin{equation}
 H_{0}=\frac 12 z \mu_- J_{-}^{(z)}+\tau J_{0}^{(z)},
\end{equation}
can be extracted from the previous computations, since (\ref{eta}) does not depend on $\eta$, 
yielding the same trigonometric P\"oschl-Teller potential.

For $\beta \geq 1$, the potential is periodic with period $\pi$, and the eigenvalues can be written as  
\beqa
E_n & = & \frac{1}{2}\mu_- z (\alpha +\beta +2 n)^2-\frac{ \tau^2}{2 \mu_- z},\nonn
    & = & \frac{1}{2}\mu_- z (2 n+ \lambda+2 )^2+ \tau  (2 n+ \lambda+2 ),
\label{ex20}
\eeqa
while the eigenfunctions satisfying the physical boundary condition, $|\Psi_n^0(y)| \rightarrow 0$ at $y \rightarrow \pm \pi/2$, are given by
\begin{equation}
\Psi_n(y) = 
{\cal N}_n (\alpha,\beta)\sin^\mu( y)\cos^\beta( y )~ _2F_1(-n,n+\alpha+\beta,1/2+\beta,\cos^2(y )).
\end{equation}

However, the limit $\eta \rightarrow \infty$ cannot be derived from the previous results. It should be obtained by taking the limit $\eta\rightarrow\infty$ of  \eqref{veta}. Thus, the change of variable from $x$ to $u$ can be obtained as the limit of \eqref{mseta}:
\beqa
-\uni x = -\frac{1}{2 z} \ln(1- \re^{2 \uni y}),~~~ y=u \sqrt{\mu_- z },
\eeqa
and the problem is mapped to the equation
\begin{equation}
\label{IPM1}
 -\frac{d^2 \phi(y)}{dy^{2}}+ \frac {2 }{\mu_- z} \left(U(y)-  \varepsilon\right)\phi(y)=0,
\end{equation}
with
\beqa
U(y,z)& =& \left \{
\begin{array}{ll}
-\frac {\tau^2}{2 \mu_- z},& 0<y<\pi,\\
\infty , & y = 0, \pi, 
\nonn
\end{array}
\right.\nonn
U(y,z)&=U&(y+\pi).
\eeqa

Eigenvalues and eigenfunctions can be straightforwardly written as 
\begin{eqnarray}
\varepsilon_n & = & \frac 12 \mu_- z ~( n)^2-\frac{\tau^2}{2 \mu_- z}, \nonn
\phi_n(y)& = &  \frac{1}{\sqrt{\pi}}\sin(n y),~n \in \mathbb{Z}. 
\end{eqnarray}

Again, we can compare these results with the ones for the finite-dimensional representations, whose eigenvalues for~\eqref{hj02} are~\cite{nos24}:
\beqa
E_{\pm~m}^{FD}= \frac 12 \mu_- z ~m^2 \pm m~\tau,~~~\tau=\sqrt{\mu_0^2+ 2 \mu_+ \mu_-},
\label{ex2fd}
\eeqa
with $m=2 k +1,~(k=0,...,d/2)$ for the case of a representation with even dimension d, and $m=2 k,~( k=0,...,(d-1)/2) $ for the case with odd dimension d.

Therefore, it can be observed that letting $(2 n+ \lambda + 2)$ to take the values $m \in \mathbb{N}$, the infinite-dimensional eigenvalues of \eqref{ex20} coincide with $E_{+~m}^{FD}$ (\ref{ex2fd}) of the finite-dimensional representation. 

Also, note that to leading order in $z$, the eigenvalues of the infinite and the finite-dimensional representations behave as $E_q\rightarrow \frac 12 \mu_- z~(2 q)^2$. Finally, if $\text{sign}(\mu_-) \neq \text{sign}(\mu_+)$, $\tau$ can take complex pair-conjugate values. 

\subsection{Example 3}\label{sec33}

Indeed, more involved choices of the functions $(\mu_0(x),\mu_\pm(x))$ can lead to the definition of new models on $U_{z}(sl(2, \mathbb R))$.For instance, let us consider the generic Hamiltonian of \eqref{hamil} with
\begin{eqnarray}
\mu_0(x)&=& \mu_0, \nonn
\mu_-(x)&=& \uni \mu_- e^{i x z} \frac{\sin (x z) \tan^2(x z)}{z^{3}},\nonn
\mu_+(x)&=& 
-\frac{\mu_+} 4  \frac{\sin^2(x z)}{z^2}.
\end{eqnarray}

This Hamiltonian can be written as a PDM system by applying the similarity transformation: 
\begin{flalign}
\Gamma(x) = 
\left ( \frac {2} {t (x, z)} - 1  \right ) {t (x, z)}^{-(\frac {\mu_ 0 } {\mu_ -} z^2 + \frac {\lambda } {2})}  
(1 - t (x, z))^{\frac {1} {4} (\lambda + 4)}
\re^{2 \frac {\mu_ 0 } {\mu_ -} z^2 \frac {(2 t (x, z) - 1)} {t (x, z)^2}}, \nonn
\end{flalign}
with the same function $t(x,z)$ given in \eqref{txz}. The transformed Hamiltonian reads
\beqa
h=\Gamma(x)\,H\, \Gamma^{-1}(x)= -\frac{1}{2}\frac{d}{dx}\left(\frac{1}{m(x,z)}\frac{d}{dx} \right)+{V}(x),
\eeqa
where the PDM function is
\begin{eqnarray}
m(x,z)&=& -\frac{z^4 \cot^2(x z) \text{csc}^2(x z)}{2 \mu_- }.
\end{eqnarray}
By performing the following change of variable
\beqa
-\uni x &=& - \frac 1 z \mbox{arcsin}\left(\frac{z}{ \sqrt{2 \mu_-} u}\right),
\eeqa
and afterwards applying the PCT method, the problem can be presented as the constant mass Schr\"odinger equation
\beqa
-\frac 12 \mu_- \frac{{\rm d}^2 \phi(y)}{{\rm{dx}}^2} + U(y,z) \phi(y) = \varepsilon \phi(y),
\eeqa
with $y=\sqrt{\mu_-} u$ and the $z$-dependent potential being
\beqa
U(y,z)&=& 
 \mu_0 \left( 3-\frac{2 z}{\sqrt{2 y^2+ z^2}} \right) + \frac{2 \mu_0^2}{\mu_-} \frac{y^2 (2 y^2+ z^2)}{\left( z+\sqrt{2 y^2+ z^2}\right)^2}+\nonn
 & &
 \frac {\mu_-} 4 \frac {  y^2 \lambda (\lambda +2 )+ z^2 (2 +2 \lambda +\lambda^2)}{\left( 2 y^2+ z^2\right)^2 } +
\frac{\mu_+} 2 \frac{(y^2+z^2)}{\left( 2 y^2+ z^2\right)^2}+ \nonn
 & &\frac{( \mu_+ -\mu_-) z^4}{8 y^2\left( 2 y^2+ z^2\right)^2}.
\label{potej3}
\eeqa

As we mentioned before, the differential realization of the algebra $\uz$ given in \eqref{dif}, makes it immediate to discern the limit as $z$ approaches zero, yielding the analogous model defined on the algebra $sl(2,\RR)$. However, the question about the behaviour of the representations and models defined on the quantum algebra $U_{z}(sl(2, \mathbb R))$ in the limit $z\to\infty$ shows intriguing features. 

It seems obvious that considering the $z\to\infty$ limit of the differential realization~\eqref{dif} proves unfeasible, as each function accompanying the derivatives diverges. Nevertheless, the transformation here presented into Hamiltonians with constant mass opens a novel path to get a deeper insight into this problem. More explicitly, in the last example we have that the $z\to\infty$ limit of the potential~\eqref{potej3} is well-defined, namely
\beqa
\lim_{z \rightarrow \infty}U(y,z)=\frac{\mu_0^2 y^2}{2 \mu_-}+\mu_0+\frac{\mu_+-\mu_-}{8  y^2} \, ,
\label{zinfty}
\eeqa
which represents a harmonic oscillator potential plus a $y^{-2}$ barrier, and the spectrum of this system is well known.

Moreover,  it is worth stressing that the undeformed $z\to 0$ limit of the potential~\eqref{potej3} has exactly the same structure as~\eqref{zinfty}, but the constants governing the term $y^{-2}$ change, namely
\beqa
\lim_{z \rightarrow 0} U(y,z)=
 3 \mu_0  + \frac{2 \mu_0^2}{\mu_-} y^2+
 \frac {(\mu_- \lambda (\lambda +2) +\mu_+)} {8 y^2} .
\eeqa
This variation becomes pivotal, as the sign of the $y^{-2}$  coefficient determines the presence of either positive or negative barriers for the harmonic oscillator term. In fact, by exploring the deformation parameter from $z=0$ to $z \rightarrow \infty$, we can observe how the types of barriers in the resulting potential change in terms of $z$. 

\begin{center}
\begin{figure}
\includegraphics[width=0.80\textwidth]{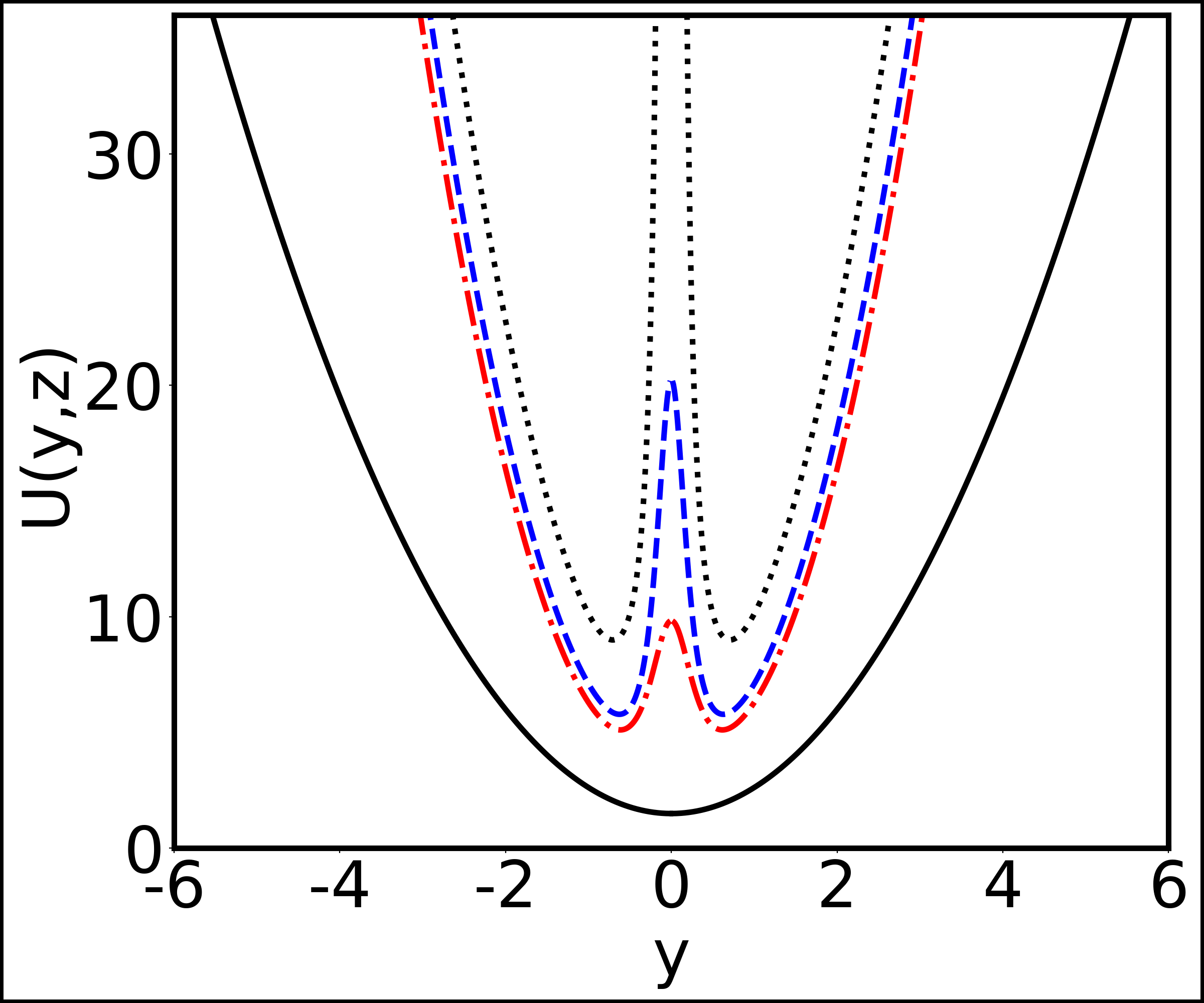}
\caption {Potential $U(y,z)$ of \eqref{potej3} as a function of $y$. We have considered the coupling constants values $\mu_0=3/2,~\mu_-=\mu_+=1$. Dashed-black, blue-, red- and solid-black curves correspond to values of $z=0, ~0.4,~0.6$,  and the limit $z \rightarrow \infty $, respectively.}
\label{fig:f5}
\end{figure}
\end{center}

In particular, for $z \neq 0$ we have that the potential of \eqref{potej3}
diverges for $y=0$, except  when $\mu_+ = \mu_-$. In Figure \ref{fig:f5}, we display the behaviour of the potential $U(y,z)$ of \eqref{potej3} as a function of $y$ in the case $\mu_0=\mu_-=\mu_+=1$. Dashed-black, blue-, red- and solid-black curves correspond to values of $z=0, ~0.25,~0.5$,  and the limit $z \rightarrow \infty $, respectively. Here it becomes evident that there exists a smooth transition, ruled by the quantum deformation parameter $z$, from an infinite double-well potential at $z=0$, passing through a finite double-well for any $z\neq 0$ and arriving at a harmonic oscillator in the limit $z\to\infty$.

\begin{center}
\begin{figure}
\includegraphics[width=0.99\textwidth]{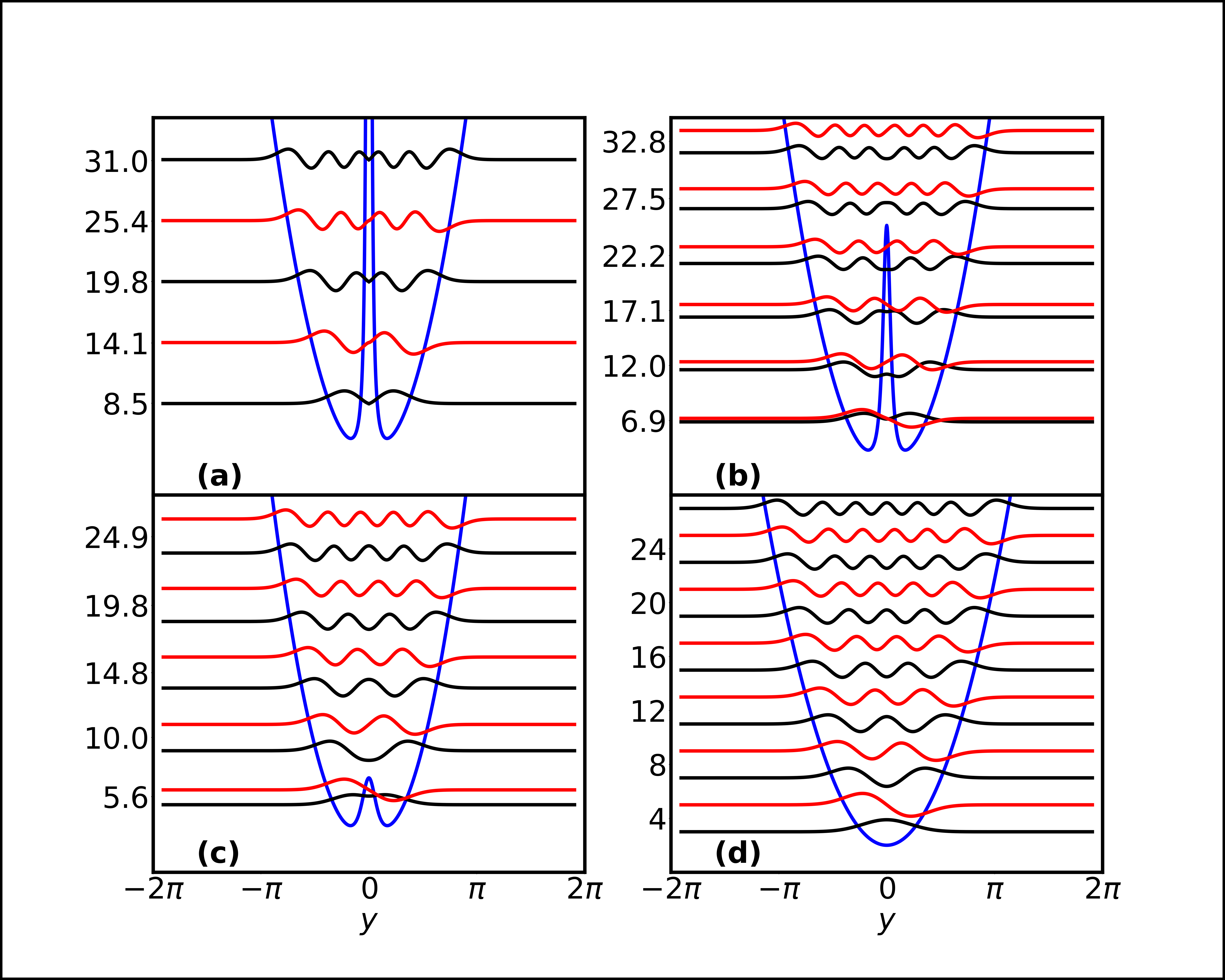}
\caption {Potential $U(y,z)$ of \eqref{potej3} as a function of $y$, and its first eigenvalues and eigenfunctions. We have considered the coupling constants values: $\mu_0=\mu_-=\mu_+=1$. Upper-left, upper-right, lower-left and lower-right Panels correspond to values of $z=0,~0.25,~0.5$,  and the limit $z \rightarrow \infty $, respectively.}
\label{fig:f6}
\end{figure} 
\end{center}

Figure \ref{fig:f6} shows again the potential $U(y,z)$ of \eqref{potej3} as a function of $y$, together with its first eigenvalues and eigenfunctions, again in the case with $\mu_0=\mu_-=\mu_+=1$. Upper-left, upper-right, lower-left and lower-right Panels correspond to values of $z=0,~0.25,~0.5$,  and the limit $z \rightarrow \infty $, respectively. Eigenvalues and the eigenfunctions have been obtained numerically.


\subsection{Transformations for a generic Hamiltonian}

All the previous results can be set in a more general framework. In fact, given the generic family of Hamiltonians
\begin{equation}
H = \mu_-(x) J_{-}^{(z)}+\mu_0(x) J_{0}^{(z)}+\mu_+(x),
\end{equation}
we can introduce the operator
\begin{flalign}
\Gamma(x)= {\cal C} \left(\frac{\sin (x z)}{\sin(z)}\right)^{-\frac{\lambda +1}{2}} 
\re^ {-\int_1^x \frac{\mu_-'(t)-2 \uni \mu_0(t)}{2 \mu_-(t)} \, dt+\frac{1}{2} \uni (x-1) z},
\label{transG}
\end{flalign}
to construct a new Hamiltonian, $h$, through a similarity transformation
\begin{equation}
h  =\Gamma(x) H \Gamma^{-1}(x)= -\frac{1}{2}\frac{d}{dx}\left(\frac{1}{m(x,z)}\frac{d}{dx} \right)+V(x,z),
\end{equation}
where the PDM function is
\begin{eqnarray}\label{zpt}
m(x,z)= -\frac{\uni z e^{\uni x z} \csc (x z)}{2 \mu_-(x)},
\end{eqnarray}
and the potential reads
\begin{eqnarray}
V(x,z) &=& \mu_+(x)+\frac{\sin ^2(x z) \left(2 \mu_0(x)+i \mu_-'(x)\right)^2}{2 z \mu_-(x) \left(e^{2 i x z}-1\right)}- \nonn
&&\frac{(\lambda  (\lambda +2)+2) z^2 \mu_-(x)+(\cos (2 x z)-1) \left(\mu_-''(x)-2 i \mu_0'(x)\right)}{2 z \left(e^{2 i x z}-1\right)}+ \nonn
&+&\frac{e^{-i x z}
  \mu_-'(x) \sin (x z)}{e^{2 i x z}-1}+\frac{z \mu_-(x) e^{-2 i x z}}{2 \left(e^{2 i x z}-1\right)}
\end{eqnarray}
Finally, by performing the change of variables $u=\int \sqrt{ m(x)} dx $ and by finally introducing the transformation $W(u)=\re^{w(u)} q(u)$, we end up with the Schr\"odinger problem
\beqa
\mathfrak{h}= -\frac 1 2 \frac {{\rm d}^2}{{\rm d u}^2}+ U(u)\, ,
\eeqa
where the potential $U(u)$ is given by~\eqref{potentialU}. 



\section{Physical applications.}\label{appli}

The generalization presented in the last Section allows the construction of specific Hamiltonians that can provide effective models for physical systems of interest. In particular, we shall study a particular case of the Hamiltonian of \eqref{hamil} with:
\begin{flalign}
\mu_-(J_+^{(z)}) =& ~g,\nonn
\mu_0(J_+^{(z)}) =& ~0,\nonn
\mu_+(J_+^{(z)}) =& ~c_1 ~ \re^{- z J_+^{(z)}}-c_2 ~ \re^{-2  z J_+^{(z)}}+\frac{c_3}{2} ~ \text{csch}( z J_+^{(z)})+
 \frac{c_4}{2} ~ ( \coth ( z J_+^{(z)})-1) \nonn
&+ \frac{g}{4}~  z \left(e^{2  z J_+^{(z)}}-4-\lambda (\lambda +2)\right) (\coth ( z J_+^{(z)})+1)
\label{fmus}
\end{flalign}

Through the transformation of \eqref{transG}, which in this case reads
\beqa
\Gamma(x)=
\left(\frac{\sin (x z)}{\sin(z)}\right)^{-\frac{\lambda +1}{2}} \re^{\frac{1}{2} \uni (x-1) z},
\eeqa
the Hamiltonian can be transformed into a PDM system
\beqa
h=\Gamma(x) H \Gamma(x)^{-1}= -\frac{1}{2} \frac{d}{dx}\frac{1}{m(x,z)} \frac{d}{dx}+V(x,z),
\eeqa
with mass function given by
\beqa
m(x,z)=\frac{z}{2 g} (1-\uni ~ \cot (x z)),
\eeqa
and whose associated potential reads
\begin{flalign}
V(x,z)= c_1 \re^{ \uni x z} -c_2 \re^{2\uni x z}
+ \frac{ c_4~\re^{2 \uni x z}+c_3  ~\re^{\uni x z}-g}{1-\re^{2 \uni x z}  } \, .
\end{flalign}

We now introduce the change of coordinates
\beqa
u= -\frac{\uni}{\sqrt{ g ~z }} {\rm{arccoth}} \left(\frac{1+\uni}{\sqrt{\cot (x z)+\uni}}\right),
\eeqa
and by applying the PCT method, we end up with the constant mass Hamiltonian 
\begin{flalign}
& {\mathfrak h}  =  -\frac{1}{2} \frac{d^{2}}{du^{2}}+U(u,z),\nonn
& U(u,z) =  
{c_4}~ \tan ^2\left(u \sqrt{ g ~z}\right)+
{c_1}~ \sin \left(u \sqrt{ g \,z}\right)- {c_2} \sin^{2} \left(u \sqrt{ g \,z} \right)+
{c_3}~ \frac{\sin \left(u \sqrt{ g \,z}\right)}{\cos\left(u \sqrt{ g \,z}\right)^{2}} \, .
\nonn
\label{hdwell}
\end{flalign}
Now we introduce the new variable $y= \sqrt{ g \,z} \, u$, so that the Schr\"odinger equation reads
\beqa
-\frac{1}{2} \psi''(y)+ u(y)\psi(y)= \varepsilon ~\psi(y), \qquad E = g \,z \,\varepsilon \, ,
\label{seq1}
\eeqa
with
\begin{flalign}
u(y)=\frac{d_s}{2}   \tan ^2\left(y\right)-
\frac{a_s}{2} \sin \left(y\right)- 
\frac {b_s} {2} \sin^{2} \left(y \right)+
\frac {c_s} {2}~ \frac{\sin \left(y\right)}{\cos\left(y\right)^{2}},
\label{potdef}
\end{flalign}
where we have defined 
\beqa
a_s = 2 \frac{c_1}{ g ~z},~b_s=2 \frac{c_2}{ g ~z},~ c_s=2~\frac{c_3}{ g ~z},~d_s=2~\frac{c_4}{ g ~z},
\label{coefz}
\eeqa
and $y$ belongs to the interval $(-\pi/2,\pi/2)$, and $u(y)\rightarrow \infty$ at $y=\pm \pi/2$.

It turns out that the family of potentials of \eqref{potdef} are Double-Well Trigonometric (DWT) potentials, which have been extensively used in order
to model specific molecular physics phenomena~\cite{sitnitsky2017a,sitnitsky2017b,sitnitsky2018,sitnitsky2019,
sitnitsky2020,sitnitsky2021,sitnitsky2023}, since the Schr\"odinger equation given in \eqref{seq1} can be solved analytically. The reliability of the results obtained with DWT potentials has been confirmed in modelling vibrational phenomena such as ring-puckering in cyclic molecules (e.g., 1,3-dioxole and 2,3-dihydrofuran) \cite{sitnitsky2018}, as well as describing the inversion of the nitrogen atom in ammonia (NH3) \cite{sitnitsky2017a}. In this case, the deformation parameter $z$ modifies the width of the well by adjusting the period of the potential. 

\begin{center}
\begin{figure}
\includegraphics[width=0.99\textwidth]{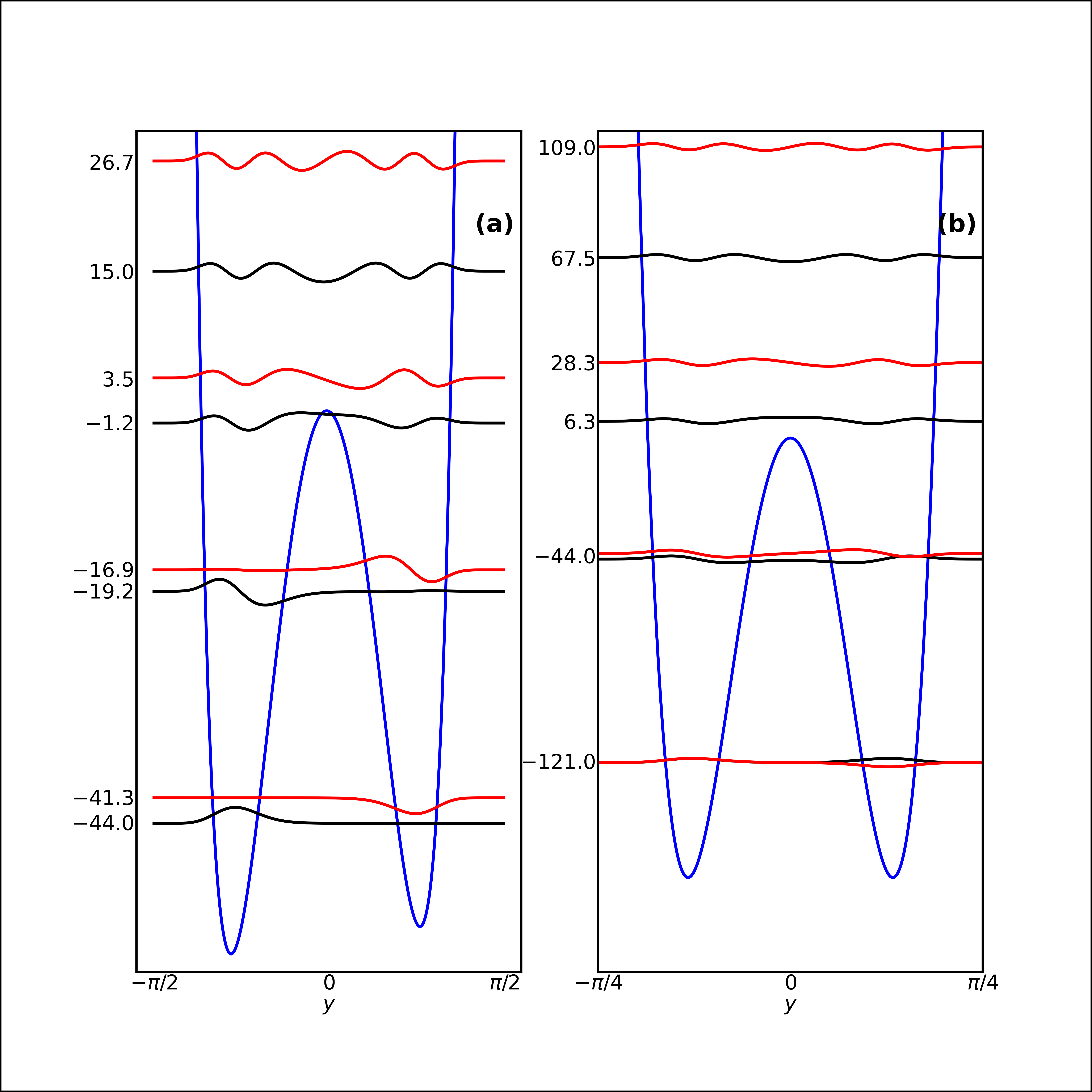}
\caption {Potential $u(y)$ of \eqref{potdef} as a function of $y$, and its first eigenvalues and eigenfunctions. The results shown in Panel (a) has been obtained 
for $a_s=23,~ b_s=360,~d_s=70$ and $c_s=2 \sqrt{35}$, in natural units \cite{sitnitsky2017a}. For Panel (b), we have considered $a_s=0,~ b_s=110^2,~d_s=2(65^2-1/4)$ \cite{sitnitsky2020}.}
\label{fig:f7}
\end{figure}
\end{center}

As reported in\cite{sitnitsky2017a}, for $c_s=\sqrt{d_s}$, the analytical eigenfunctions of \eqref{seq1} which fulfil the physical boundary condition,  are given by
\begin{flalign}
\psi(y)= {\cal N} \re^{-\sqrt{b_s} \sin(y)}
\left( \tan \left( \frac \pi 4 + \frac y 2 \right)\right)^{\sqrt{d_s}}  
{\rm {HeunC}} \left(\alpha,\beta,\gamma, \delta,\eta ;\frac{1+ \sin(y)} 2\right),
\end{flalign}
with 
\begin{flalign}
\alpha=- 4 \sqrt{b_s}, ~\beta=-\frac 12 +\sqrt{d_s},~\gamma=-\left(\frac 12 + \sqrt{d_s} \right), ~\delta= 2 a_s, ~\eta=\frac 38 -a_s-b_s-\frac{d_s}2 -\varepsilon, 
\end{flalign}
and here ${\rm {HeunC}} \left(\alpha,\beta,\gamma, \delta,\eta ;q \right)$ is the confluent Heun function. The corresponding eigenvalues are obtained by imposing the boundary condition, which is $|\psi(y)|\rightarrow 0$ as $y \rightarrow \pm \pi/2$, and thus
\beqa
{\rm {HeunC}} \left(\alpha,\beta,\gamma, \delta,\eta ;1\right)=0.
\eeqa
In the work \cite{sitnitsky2017a} the energy levels for the hydrogen bound in KHCO$_3$ cristal \cite{KHCO3} are reproduced by assuming particular values of the parameters $a_s,~b_s$ and $d_s$. We plot, in Figure \ref{fig:f7} (a) the potential, the first eigenvalues and their correspondent eigenfunctions for $a_s=23,~ b_s=360,~d_s=70$ and $c_s=2 \sqrt{35}$, in natural units.


Another example of physical interest is the DWT symmetric potential obtained by taking 
$a_s=c_{s}=0$ and $4 d_s +1= 4 m^2$, with $m \in \NN$ \cite{sitnitsky2020}. In this case the eigenfunctions are given by
\begin{equation}
\label{eq:psiq}
\phi_{n,m}(b_s,q) \;=\; \cos(q)^{\tfrac{1}{2}}\;\bar{S}_{m (n+m)}\bigl(\sqrt{b_s};\,\sin q \bigr),
\end{equation}
 and the corresponding energy eigenvalues read
\begin{equation}
\label{energylevels}
E_{n,m}(b_s) =\frac{1}{2}( \lambda_{m (n+m)} +\frac{1}{2}-m_s^{2}-b_s),
\end{equation} 
where $4 m^2= 4 d_s + 1$, with $m \in \NN$. Here
$\bar{S}_{m(n+m)}\bigl(p; s\bigr)$ 
is the normalised angular prolate spheroidal function. These functions depend on a continuous parameter $b_s$ that governs the degree of elongation, as well as two discrete parameters $n$ and $m$, which determine the order of the solution.  
The quantities $\lambda_{m(n+m)}$ denote the eigenvalues of $\bar{S}_{m(n+m)}\bigl(b_s; s\bigr)$ \cite{flammer2014spheroidal,abra}. 

In \cite{sitnitsky2020}, a suitable choice of the parameters makes it possible to compute the energy levels for the hydrogen bond in the Zundel ion H$_5$O$^+_2$ \cite{H5O2}. Figure \ref{fig:f7} (b) depicts the the potential, the first eigenvalues and their correspondent eigenfunctions for $a_s=0,~ b_s=110^2,~d_s=2(65^2-1/4)$ and $c_s=0$, in natural units.


The reader is kindly referred to \cite{sitnitsky2017a,sitnitsky2020} for further details on the choice of parameters and comparison with the experimental data.


\section{Conclusions and outlook}\label{con}

In this work, we have introduced a $\pt$-symmetric differential realisation of the $\uz$ Hopf algebra acting on an irreducible infinite-dimensional module. From it, a large family of $\pt$-symmetric quantum systems is defined through Hamiltonians~\eqref{hamil} defined on such realization of the $\uz$ Hopf algebra. 

A general method to solve all these Hamiltonians is proposed in two steps. Firstly, we have shown how to map the original problem into a PDM Schr\"odinger equation by performing a suitable similarity transformation. Secondly, the PCT formalism is used to transform the PDM Hamiltonian into a fully equivalent constant mass Schr\"odinger equation with a potential that depends on the quantum deformation parameter $z$. The latter constant-mass problem can be then exactly solved for some choices of the initial Hamiltonian, and in any case the final system encodes the essential features of the original $\pt$-symmetric Hamiltonian, thus allowing the analysis of the role played by $z$ within the model.

By following this apporach, we have studied several specific Hamiltonians among the family~\eqref{hamil}. For all of them we have analysed the regions in the model space of parameters where the spectrum corresponds to bound states, which is the $\pt$-symetric phase. When possible, we have presented analytical results; otherwise, we have computed the eigenvalues and eigenfunctions numerically. 

In particular, we have started by analysing an undeformed $sl(2,\RR)$ Hamiltonian, in order to provide an overview of the essentials of the formalism. Its counterpart in terms of the differential representation of $\uz$, has been discussed in Subsection \ref{sec32}.  In Subsection \ref{sec34}, we have studied the system arising from the differential representation of the Hamiltonian given in \eqref{hj02}, which we had previously analysed in \cite{nos24} in its finite-dimensional representation. Afterwards, a particular choice of functions for the Hamiltonian (\ref{hamil}) has been found in order to construct effective two-well potentials in Subsection \ref{sec33}, where the parameter $z$ is shown to control the shape of the potential: the $z\to 0$ case provides an infinite barrier, which is transformed into a double well when $z\neq 0$, and the limit $z\to\infty$ leads to the harmonic oscillator potential. Remarkably enough, in Section \ref{appli} it is shown that the Hamiltonian of \eqref{hamil}, with $(\mu_0,\mu_\pm)$ functions given by \eqref{fmus}, turns out to be equivalent to a DWT potential, and can be used to describe realistic physical problems as the inversion of the nitrogen atom in ammonia (NH3) \cite{H5O2} or the levels for the hydrogen bond in KHCO$_3$ cristal \cite{KHCO3}. 

Summarizing, we have demonstrated that Hamiltonians of the type \eqref{hamil}, when written in terms of the differential $\pt$-symmetric realisation of the non-standard $\uz$ algebra, can be transformed into PDM Hamiltonians through suitable similarity transformations, and can be further converted via PCT transformations in order to model the bound spectrum of double-well systems and barriers, e.g. general P\"oschl-Teller or DWT potentials.  Work is in progress concerning the analysis, in the model space of parameters for $\uz$ Hamiltonians, of the appearance of Exceptional Points and the study of non-$\pt$ symmetric phases.


\section*{Acknowledgements}

The authors acknowledge partial support from the grant PID2023-148373NB-I00 funded by MCIN/AEI/ 10.13039/501100011033/FEDER -- UE, and the Q-CAYLE Project funded by the Regional Government of Castilla y Le\'on (Junta de Castilla y Le\'on) and the Ministry of Science and Innovation MICIN through NextGenerationEU (PRTR C17.I1). M.R. is grateful to the Universidad de Burgos for its hospitality. M.R. and R.R. have been partially supported by grants 11/X982 of the University of La Plata (Argentine) and PIP0457 CONICET.

\section*{References}

\providecommand{\newblock}{}

\end{document}